# Nanoscale Terahertz Conductivity and Ultrafast Dynamics of Terahertz Plasmons in Periodic Arrays of Epitaxial Graphene Nanoribbons

*Arvind Singh[1], Hynek Němec[1], Jan Kunc[2], and Petr Kužel[1*]*

[1] *Institute of Physics of the Czech Academy of Sciences, Na Slovance 2*
*18221 Prague 8, Czech Republic*

[2] *Faculty of Mathematics and Physics, Charles University, Ke Karlovu 3,*
*12116 Prague 2, Czech Republic*

[*] kuzelp@fzu.cz

**Abstract:** Dynamics of plasmons in nanoribbons of (hydrogen intercalated) quasi-free-standing single layer graphene is studied by terahertz spectroscopy both in the steady state and upon photoexcitation by an ultrashort near infrared laser pulse. The use of two-dimensional frequency domain analysis of the optical pump – THz probe signals allows us to determine the evolution of carrier temperature and plasmon characteristics with ~100 fs time resolution. Namely, we find that the carrier temperature decreases from more than 5000 K to the lattice temperature within about 7 ps and that during this evolution the carrier mobility remains practically constant. The time-resolved THz conductivity spectra suggest that graphene nanoribbons contain defects which act as low potential barriers causing a weak localization of charges; the potential barriers are overcome upon photoexcitation. Furthermore, the edges of graphene nanoribbons are found to slightly enhance the scattering of carriers. The results are supported by complementary measurements using THz scanning near-field microscopy which confirm a high uniformity of the THz conductivity across the sample and demonstrate high enough sensitivity to resolve even the impact of nanometric terrace steps on SiC substrate under the graphene monolayer.

## 1. Introduction

Surface plasmons in subwavelength structures of graphene are collective oscillations of Dirac quasiparticles [1,2,3] which are essentially electromagnetic waves propagating inside and in the close vicinity of the graphene [4,5,6,7]. The confinement of these waves strongly enhances the light-matter interaction, thus increasing the linear absorption and allowing exploitation of nonlinear phenomena already at moderate intensities in atomically thin subwavelength devices [8,9,10,11,12,13]. Graphene versatility together with the tunability of the plasmons makes graphene structures excellent candidates for next-generation plasmonic devices [14,15,16,17,18,19,20,21,22] with applications in ultrafast electronics, imaging and quantum technologies from the terahertz to the infrared regime. The unique properties of the graphene sheets such as high carrier mobility at room temperature and electrical, optical and chemical tunability of their opto-electronic properties ascribe to graphene plasmons a great potential especially for terahertz detection and sensing applications [23] in terms of the interaction strength, tunability and speed of the response. All these applications rely on a deep understanding of the optoelectronic properties of nanostructured graphene, which are determined by ultrafast relaxation dynamics of the hot carriers.

Plasmons in graphene have been mostly observed in micron-scale ribbons where the resonance frequency is either controlled via ribbon parameters or by electronic gating. Due to the exceptionally small electronic specific heat capacity of graphene, hot-carrier effects are pronounced and they influence the nonlinear dynamics of plasmons in graphene structures, which may be important for potential applications of graphene nanostructures in ultrafast plasmonic devices [8,9,10]. These plasmonic nonlinearities were investigated recently using transient THz transmittance induced by intense narrow-band terahertz (THz) pulses [8,9,10].

In this paper, we study the THz response of graphene ribbons excited by 800 nm femtosecond laser pulses. To retrieve the most complete dynamical picture of the carrier evolution in the photoexcited graphene nanoribbons we combined analyses using both a conventional approach applicable for a slow dynamics [24] and a two-dimensional (2D) frequency-domain approach for characterization of sub-picosecond dynamics [25,26,27,28]. Comparison of the response for both polarizations of the THz probe with respect to the nanoribbon direction is important to elucidate the role of growth defects and the graphene ribbon edges. As a complementary experiment we applied scattering-type near field THz microscopy (THz-SNOM) which provides information about the local conductivity on ~100 nm scale and simultaneously records also the AFM topographical image of the surface.

## 2. Experimental details

### 2.1. Steady-state setup and sheet conductivity measurements

The THz steady-state conductivity spectra of the graphene ribbons were measured in transmission configuration using a conventional femtosecond oscillator-based time-domain THz spectroscopy setup basically described in [29].

In the case of a two-dimensional (2D) material like graphene on a substrate, the THz sheet conductivity $\sigma$ can be retrieved from the experimental data using the Tinkham formula [30] adapted for phase sensitive measurements:

$$\frac{E^0(\omega)}{E^0_{ref}(\omega)} = \frac{N_s + 1}{N_s + 1 + z_0 \sigma} e^{i\omega(N_s-1)(d_s-d_r)/c}, \quad (1)$$

where $z_0$ is the vacuum wave impedance. $E^0(\omega)$ is the Fourier transformation of a sample waveform (measured with graphene film on substrate with thickness $d_s$ and refractive index $N_s$) and $E^0_{ref}(\omega)$ is the Fourier transformation of a reference waveform (measured with a bare substrate with thickness $d_r$). The complex exponential term in Eq. (1) accounts for the difference between phase shifts induced by the reference and sample substrate, which has to be carefully accounted for when calculating the conductivity of thin films. Therefore, to ensure the most accurate data, we systematically optimized the value of the substrate thickness difference $d_s - d_r$ by comparing the sheet conductivity spectra obtained from the directly transmitted THz pulses [$E^0(\omega)$ and $E^0_{ref}(\omega)$] and from the first Fabry-Pérot reflections within the substrates, as explained in the Supplementary material of our previous study [24].

### 2.2. Optical pump – THz probe setup and transient THz conductivity measurements

The complex transient transmittance spectra of the studied samples were measured in optical pump – (multi-)THz probe setups powered by a Ti:sapphire ultrafast amplified laser system (Spitfire ACE, 5 kHz repetition rate, 40 fs pulse length), see Fig. 1. The fundamental laser wavelength of 800 nm was used for the sample photoexcitation.

For the optical pump – THz probe spectroscopy the generation and detection occur in identical (110)-oriented 1 mm thick ZnTe crystals. To achieve the collinear pump beam geometry, we use a 1 mm thick and 2 inches in diameter fused silica plate introduced into the THz beam path under 45°. The Fresnel losses of the plate for the *p*-polarized THz pulse are minimal since the angle is close to the THz Brewster angle (more than 90% of the THz field is transmitted). The plate also features a high-reflective dielectric coating on its front side for the 800 nm excitation pulses to allow an efficient introduction of the optical pump pulse with a large diameter. The setup is thus convenient to reach the ultimate time resolution and ensure a homogeneous illumination of the sample by the pump beam. The pump beam chopper is synchronized with the laser output (chopper CH2, 2.5 kHz), the THz beam is modulated at an incommensurate frequency (chopper CH1, 1.127 kHz).

In optical pump – multi-THz probe experiments [31] linearly polarized multi-THz pulses are generated by two-color plasma mixing in the air [32]. The residual power of the optical pump beam is filtered by a silicon plate oriented close to the Brewster angle. The pump beam impinges on the sample under an angle of 7° with respect to the probe beam; since the diameter of the multi-THz beam spot is smaller than 0.4 mm for the whole frequency content, the experimental time resolution is better than 200 fs. The detection is performed by using the air-biased-coherent-detection (ABCD) technique [33]: the multi-THz beam is focused collinearly with the sampling beam to a spot between two 1-mm-thick electrodes with a gap of 1 mm between them. The electrodes are supplied with ±1.5 kV rectangular pulses at 500 Hz (synchronized with the laser repetition rate) which superimpose to the THz field. The second harmonic beam generated in the detection process is sent through a set of 400 nm band-pass filters into an avalanche photodiode. The pump beam chopper is modulated at a frequency incommensurate with the laser repetition rate (chopper CH2, 2.126 kHz). The details and further characteristics of our multi-THz setup are provided in [31].

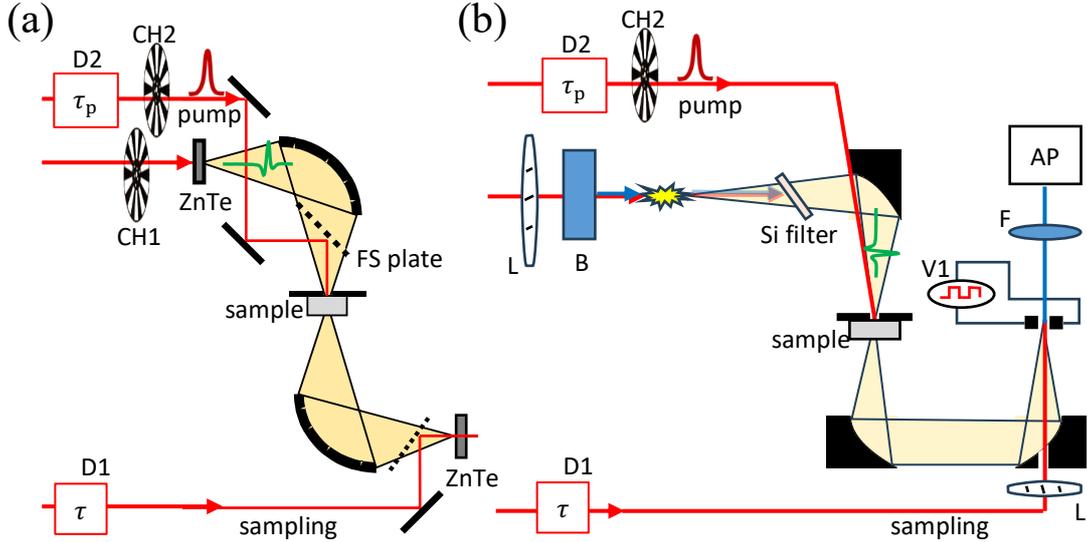

Figure 1. Experimental setups for time-resolved (a) THz and (b) multi-THz spectroscopies. D1, D2: optical delay lines, CH1, CH2 choppers, L: converging lens, B: assembly of BBO (second harmonic generation), $\alpha$-BBO (time plate controlling the phase between the 800- and 400-nm pulse) and dual 800/400 nm waveplate; AP: avalanche photodiode, F: 400 nm bandpass filters; V1: ±1.5 kV voltage source at 500 Hz synchronized with the laser.

In both experimental setups the sampling of THz (multi-THz) pulse is controlled by the delay line D1 (time $\tau$) and the pump-probe delay $\tau_\text{p}$ is controlled by D2. The transient signal $\Delta E(\tau, \tau_\text{p})$ and the reference transmission $E^0(\tau)$ are then obtained at the same time by a double demodulation technique.

The measured sheet photoconductivity $\Delta\sigma$ is calculated from the transient ($\Delta E$) and reference ($E^0$) signals in the thin film approximation and in the small-signal limit ($\Delta E \ll E^0$) as follows [34]:

$$\Delta\sigma(\omega) = -\frac{1 + N_s}{z_0} \frac{\Delta E(\omega)}{E^0(\omega)}. \qquad (2)$$

The total measured sheet photoconductivity is given by a sum of the following terms:

$$\Delta\sigma(\omega, \tau_p) = \sigma_e(\omega, \tau_p) - \sigma(\omega) + \Delta\sigma_{SiC}(\omega, \tau_p), \qquad (3)$$

where $\tau_p$ is the pump-probe delay, $\sigma_e$ is the sheet conductivity of photoexcited graphene layer and $\sigma$ is its ground state (steady-state) sheet conductivity. The photoconductivity of graphene (the first two right-hand-side terms) dominates; however, Eq. (3) takes into account also the observed small photoconductivity $\Delta\sigma_{SiC}$ of the SiC substrate [24]; the spectra $\Delta\sigma_{SiC}(\omega, \tau_p)$ were independently measured using a bare substrate.

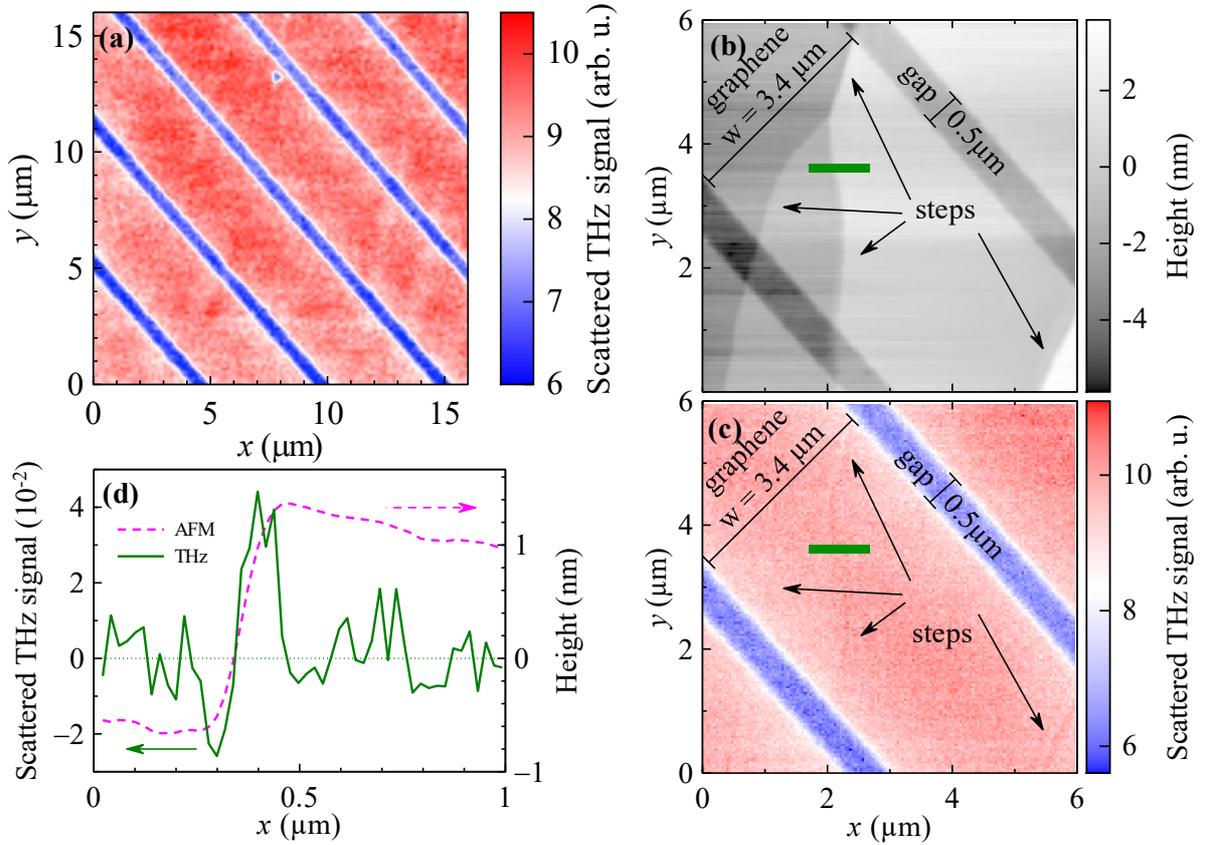

Figure 2. (a) Spatial mapping of the scattered THz signal measured using the THz-SNOM device; image 16×16 μm². The width of the ribbons is $w = 3.4$ μm and they are separated by empty gaps 0.5 μm wide; the period of the array is $L = 3.9$ μm; the graphene filling fraction is 87%. (b) AFM height profile of the studied array of graphene ribbons (higher-resolution image of $5 \times 5$ μm²); the height changes due to SiC terrace steps are clearly observed. (c) Higher-resolution THz-SNOM image of the same area. In this view we also distinguish the SiC terrace steps within the graphene ribbons. (d) Comparison between the sample height recorded by AFM and the scattered THz amplitude obtained during a scan along the green horizontal line indicated in panels (b,c). For the scattered THz signal, a background (straight line) was subtracted; the level of the subtracted background is ~9, the plotted THz signal magnitude in (d) represents a change with respect to this value using the same scale.

### 2.3. THz-SNOM

The THz-SNOM setup (Neaspec) consists of an atomic force microscope, where a metallic tip (diameter ~ 50 nm) is illuminated by focused picosecond THz pulses covering spectral range 0.5 – 2.0 THz. The scattered THz amplitude, resulting from the near-field interaction between the tip and the sample surface, allows for the acquisition of THz near-field images with nanoscale resolution across the scanned region of the sample; the data presented in this paper display the THz scattered amplitude demodulated at the second harmonic of the tip tapping frequency. The AFM topography of the surface is recorded simultaneously.

### 2.4. Sample preparation

Single layer of quasi-free-standing single layer graphene (QFSLG) was grown on the Si side of a 6H-SiC substrate using the procedure described in [24,35,36]. Subsequently, standard electron beam lithography followed by oxygen plasma etching was employed to create graphene ribbons 3.4 μm wide separated by 0.5 μm gaps over an area of 5×5 mm$^2$.

## 3. Results and discussion

### 3.1. THz near field imaging

In the initial step, we examined the topography and local THz conductivity of the sample by using the THz scattering near-field microscope (THz-SNOM), Fig. 2. While vertical terraces of the SiC substrate are very clearly visible on the AFM image, Fig. 2(b), no other topographic defects in the structure are seen in these images. The sample presents excellent homogeneity from the point of view of the local THz conductivity, Figs. 2(a,c). Notice that the scattered THz signal exhibits a slight variation in the close vicinity of the nanometer-sized vertical terrace steps on the SiC substrate such that these defects are observed in the local conductivity image, Fig. 2(c), and even better, in a scan of the line of interest, Fig. 2(d). On the one hand, one can expect that the substrate edge can cause an enhanced carrier scattering in the graphene layer due to structural defects and layer discontinuities accompanied by a drop in the local carrier mobility. On the other hand, ballistic transport of carriers (i.e., enhanced local mobility) has been observed at zig-zag edge sidewall ribbons grown on 6H-SiC substrate [37,38]. In addition, the intricate structure of the graphene layer on the terrace step, characterized by variations in its slope and step height, leads to a pronounced decrease in electrostatic potential at the step boundary. This, in turn, gives rise to the formation of residual resistivity dipoles in the graphene close to the step, as documented by Wang et al. [39] and Krebs et al. [40]. The accumulation and depletion of charges at the opposite sides of the step then may trigger the movement of additional charges from the surrounding graphene, resulting in an enhanced mobility across the step. In the current state of understanding we are not able to fully interpret the observed shape of the local THz signal in Fig. 2(d) but some interplay between the above mentioned phenomena can presumably be responible for the observed bipolar character of its spatial dependence across the step shown in Fig. 2(d).

### 3.2. Steady-state terahertz conductivity and picosecond dynamics

*Experimental data and models*

Steady-state THz sheet conductivity spectra as well as transient THz sheet conductivity spectra of the patterned graphene layer were measured for both THz probing polarizations. The experimental data are shown in Fig. 3 (for $E_{\text{THz}} \perp$ ribbons) and Fig. 4 (for $E_{\text{THz}} \parallel$ ribbons). The spectral response is significantly different for the two polarizations. This is understandable since the depolarization fields developed in the graphene layer are quite different in both cases: they practically vanish for the parallel geometry while they significantly contribute and lead to a formation of the plasmon in the perpendicular geometry.

For incident THz electric field perpendicular to the ribbons, the spectral response shown in Fig. 3 reflects the separation and screening of carriers inside the ribbons, i.e., the conductivity both in steady and photoexcited state features a pronounced plasmonic oscillation resulting in a prominent conductivity peak which can be described by a Lorentz oscillator [8]

$$\sigma_\perp(\omega, T_c) = \frac{w}{L} \frac{D}{\pi} \frac{i\omega}{\omega^2 - \omega_0^2 + \frac{i\omega}{\tau_{s,\perp}}} \qquad (4)$$

where $w/L = 87\%$ is the surface coverage by the graphene ribbons (filling fraction of the graphene component). The response is controlled by the Drude weight $D$ [41], the geometry of nanoribbons—both determining the plasmonic resonance (angular) frequency $\omega_0$—and by the carrier relaxation time $\tau_s$:

$$D(T_c) = \frac{2e^2}{\hbar^2} k_B T_c \ln\left[2\cosh\left(\frac{\mu(T_c)}{2\,k_B T_c}\right)\right], \qquad (5)$$

$$\omega_0(T_c) = \sqrt{\frac{D(T_c)w}{\varepsilon_0(1 + \varepsilon_{\text{SiC}})\, L^2 \ln\left(\sec\left(\frac{\pi w}{2L}\right)\right)}}, \qquad (6)$$

$$\tau_{s,\perp} = \frac{\eta_\perp \mu(T_c)}{e_0 v_F^2}. \qquad (7)$$

Here $T_c$ and $\mu(T_c)$ denote the carrier temperature and chemical potential, respectively, $v_F \approx 10^6$ m/s is the Fermi velocity in graphene, and $\eta_\perp$ denotes the mobility of charge carriers. The scattering rate presented in the literature frequently includes a term representing LA phonon scattering which is proportional to the carrier temperature $T_c$. However, this term starts to play some role only for longer scattering times ($\gg 100$ fs) and we thus neglect it in our work. It was assumed in the previous publications [8,9,24] that the carrier mobility is independent of the carrier temperature and the Fermi level position. We believe that the scattering time is the primary parameter, which is actually measured, and which reflects the character of the motion of carriers. It means that the mobility can, in principle, vary with the pump-probe delay under the conditions of intense optical excitation and strong carrier heating. In this paper we follow the above ansatz, and we deduce the behavior of the carrier mobility upon photoexcitation from the experiment.

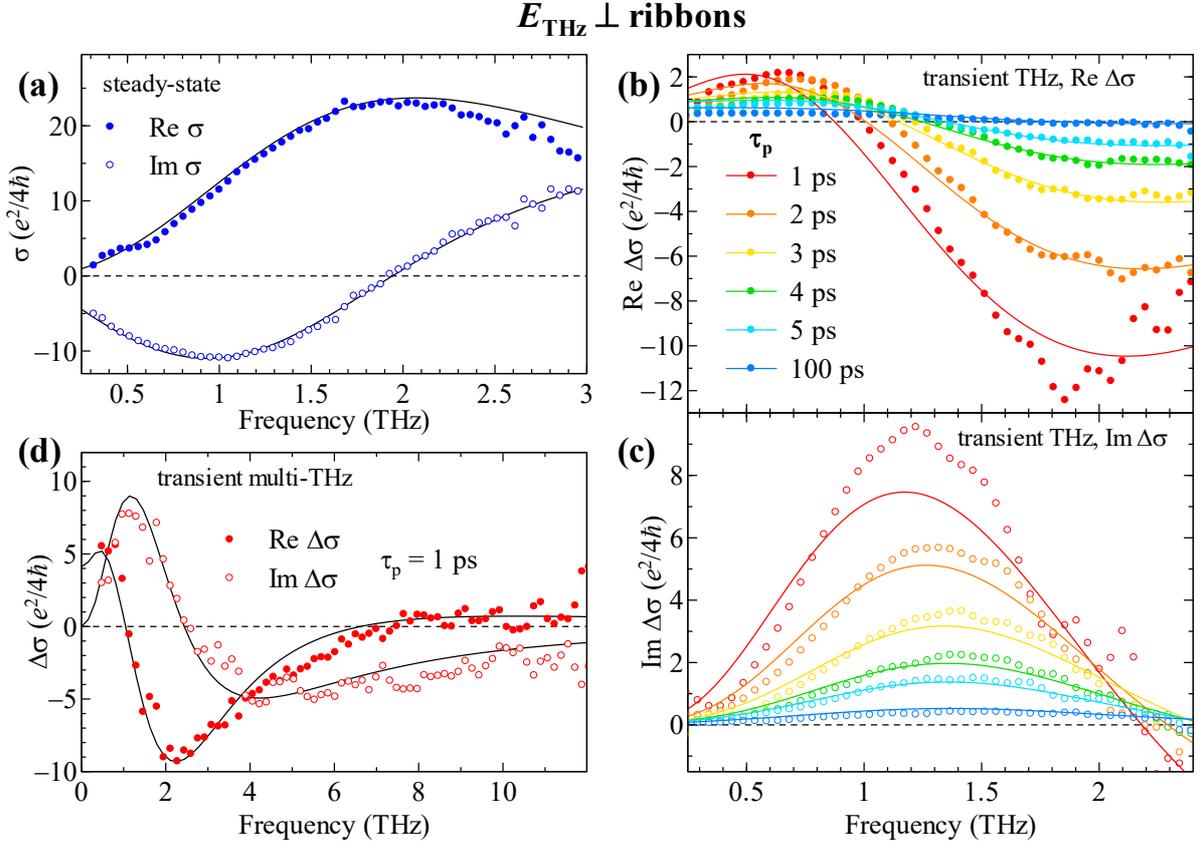

**Figure 3.** Steady-state (a) and transient (b,c,d) spectra of the complex sheet conductivity of graphene nanoribbons for THz electric field perpendicular to the ribbons. (b,c) Real and imaginary sheet photoconductivity, respectively, measured at various pump-probe delays using THz transient spectroscopy with an incident pump fluence: $F = 0.05$ mJ cm$^{-2}$ (i.e., photon fluence of $\Phi = 2 \times 10^{12}$ cm$^{-2}$ absorbed in the graphene layer); (d) real and imaginary sheet photoconductivity measured by transient multi-THz spectroscopy for $\tau_p = 1$ ps; absorbed photon fluence $\Phi = 2.3 \times 10^{13}$ cm$^{-2}$. Symbols: measured data, lines: global fit based on Eqs. (3) and (4).

For the THz probing field parallel to the ribbons, the shape of the steady state spectra is very close to the Drude response [see Fig. 4(a)]. Nevertheless, the transient spectra, as shown in Fig. 4(b–d) feature some degree of carrier localization, which is neither strong enough to form a characteristic resonance in the steady-state spectra (thus ruling out the presence of a plasmon resonance), nor weak enough to permit fitting by a simple Drude model. Instead, the spectral shape for the parallel geometry perfectly follows the so-called "modified Drude-Smith" (MDS) model [42,21]. In contrast with the purely phenomenological Drude-Smith model, the MDS approach has a clear microscopic background in accounting for the drift-diffusion current which restores the thermal equilibrium in systems without translational invariance due to, e.g., energy barriers. The physical interpretation of this model was confirmed also by a semiclassical limit of the quantum model of the THz conductivity [43]. The conductivity for this geometry then reads:

$$\sigma_\parallel(\omega, T_c) = \frac{w}{L}\frac{D}{\pi}\frac{\tau_{s,\parallel}}{1 - i\omega\tau_{s,\parallel}}\left(1 + \frac{c}{1 - i\omega\theta}\right) \qquad (8)$$

where the Drude-Smith constant $c$ describes the degree of localization of charge carriers ($c = 0$ represents the Drude model of delocalized charges, $c = -1$ describes charges fully localized within a domain) and the characteristic diffusion time $\theta \sim d^2/(10 D_{\text{dif}})$ puts in relation the diffusion coefficient $D_{\text{dif}} \sim v_F^2 \tau_s / 2$ and the characteristic dimension of the carrier confinement $d$. The other symbols have the same meaning as for the perpendicular geometry, namely, the Drude weight $D$ and the scattering time $\tau_{s,\parallel}$ are given by equations analogous to (5) and (7).

We are aware that we investigate a rather fast dynamics where, for very short pump-probe delays, the leading and trailing edge of the picosecond THz pulse probe the photoexcited sample in a different state [44]. This often results in appreciable spectral artifacts [26], which are observed in this particular case as pronounced oscillations for the shortest pump-probe delays, see the spectra observed for $\tau_p = 1$ ps in Fig. 4(b,c). Interestingly, the spectra for the perpendicular geometry [Fig. 3(b,c)] are somewhat less affected than those for the parallel geometry [Fig. 4(b,c)]. This could be related to the fact that low-frequency photoconductivity components are expected to be influenced the most by these artifacts and the low-frequency photoconductivity $\Delta\sigma_\perp$ is significantly weaker than $\Delta\sigma_\parallel$.

For this reason we performed complementary ultrafast measurements with $\sim 100$ fs long probe pulses in the multi-THz range which should be free of these artifacts, as indeed observed in Figs. 3(d) and 4(d). Since the multi-THz setup suffers from a lower signal-to-noise ratio and dynamic range than the standard THz setup [31], we used for these experiments an order of magnitude higher pump pulse fluence to obtain meaningful results.

For this reason, we also focused on sub-picosecond dynamics in a separate set of experiments as described in subsection 3.3 below.

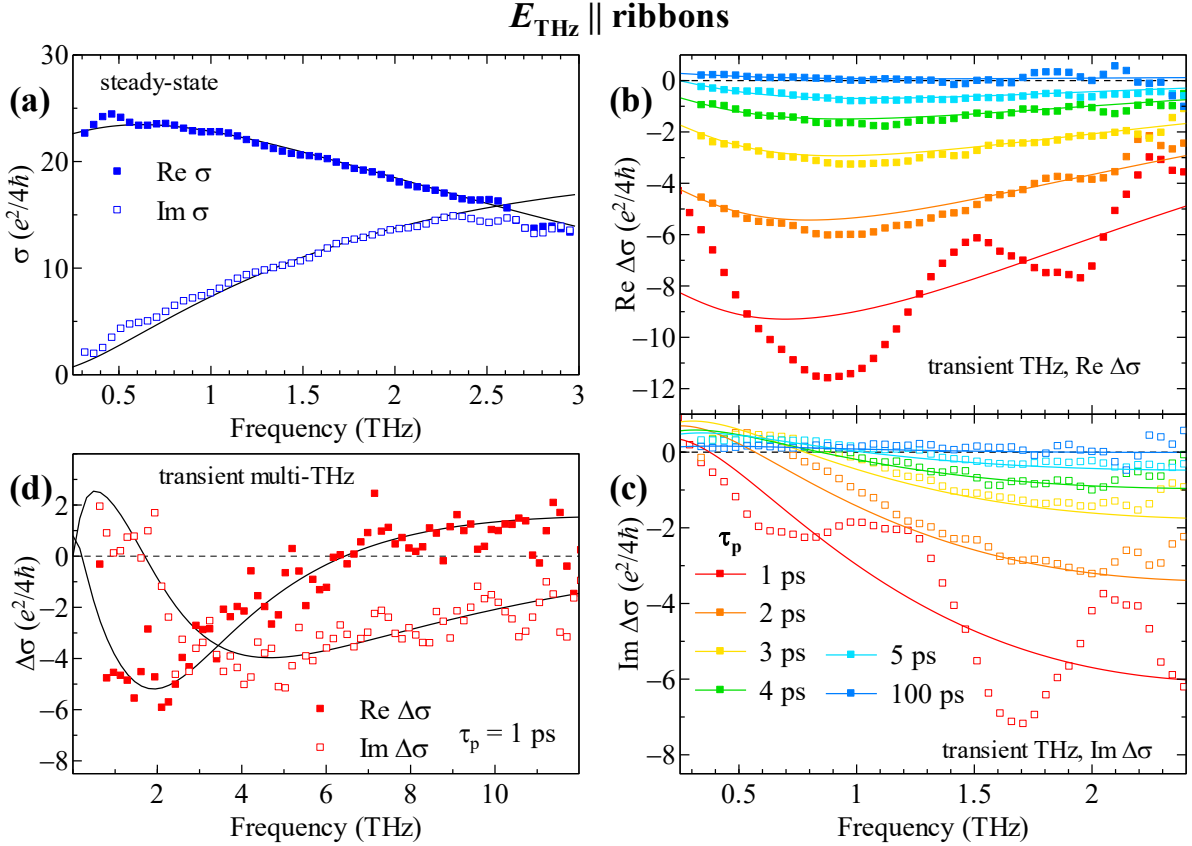

**Figure 4.** Steady-state (a) and transient (b,c,d) spectra of the complex sheet conductivity of graphene nanoribbons for THz electric field parallel to the ribbons. (b,c) Real and imaginary sheet photoconductivity, respectively, measured at various pump-probe delays using THz transient spectroscopy with an incident pump fluence: $F = 0.05$ mJ cm$^{-2}$ (i.e., photon fluence of $\Phi = 2 \times 10^{12}$ cm$^{-2}$ absorbed in the graphene layer); (d) real and imaginary sheet photoconductivity measured by transient multi-THz spectroscopy for $\tau_p = 1$ ps; absorbed photon fluence $\Phi = 2.3 \times 10^{13}$ cm$^{-2}$. Symbols: measured data, lines: global fit based on Eqs. (3) and (8).

*Fits of the data*

For the fitting of steady-state spectra we used directly Eqs. (4) or (8), depending on the geometry. For the fitting of time-resolved spectra we used the general equation (3) for the two geometries where the appropriate formula—i.e., (4) or (8)—is used both for $\sigma$ (with $T_c = 300$ K) and for $\sigma_e$ (where the carrier temperature $T_c$ is elevated and it varies with the pump-probe delay $\tau_p$). In order to maximize the meaningful physical output of the fitting we followed a specific procedure, which is described below.

We first fitted the spectra in the perpendicular geometry, which are less affected by artifacts due to the fast dynamics and therefore the fit should provide the most reliable parameters. The global fit in the perpendicular geometry was performed on all the data (both steady-state and transient) shown in Fig. 3. Each complex spectrum was parametrized by its own free value of the carrier scattering time $\tau_{s,\perp}$ and each transient spectrum had its own free value of the carrier temperature $T_c$ ($T_c$ was fixed to 300 K for the steady-state spectrum). The Fermi energy $E_F$ of the sample was introduced as a single global fitting parameter for all the spectra. The Drude weight $D$ and the plasmon resonance frequency $\omega_0$ were calculated from the value of $T_c$ using Eqs. (5) and (6), respectively. The structural parameters ($L$, $w$, $\varepsilon_{SiC}$) were kept fixed to the nominal values. The spectra of $\Delta\sigma_{SiC}(\omega, \tau_p)$ introduced in Eq. (3) were measured independently.

The spectra in the parallel geometry were measured in identical conditions (pump beam power and diameter, pump-probe delays). Therefore, for the fit of the data measured in the parallel geometry, Fig. 4, we fixed the previously obtained value of $E_F$ and also the carrier temperature values $T_c$ for each pump-probe delay. In this way, we reduce the impact of the weak artifacts caused by the ultrafast dynamics. This should be no problem from the point of view of the graphene physics since the carrier temperature should be the same under equivalent excitation conditions independently on the polarizations of the excitation and probing light. So, we were left with the following free fitting parameters: the carrier scattering time $\tau_{s,\parallel}$ and the confinement parameter $c$ for each available complex spectrum, and one global parameter—the diffusion time $\theta$—with a common value for all the spectra.

The models fit the data quite well as observed in Figs. 3 and 4. As already explained, the sheet photoconductivity spectrum obtained by THz spectroscopy at $\tau_p = 1$ ps exhibits artificial oscillations originating from the ultrafast dynamics which the fit thus cannot reproduce. In the next section we show measurements allowing a proper deconvolution of these effects in the THz range on sub-ps timescale. Note also that the complementary spectra obtained by multi-THz spectroscopy (with much shorter probing pulses, ~ 100 fs) do not exhibit such artifacts and a very good agreement with the fit up to 12 THz is obtained in both geometries [see Figs. 3(d) and 4(d)] demonstrating an overall consistency of the models. We also remark that an order of magnitude higher pump fluence used in the multi-THz experiments leads to an increase of the carrier temperature at 1 ps from 1200 K (absorbed photon fluence $\Phi = 2 \times 10^{12}$ cm$^{-2}$, THz experiments) to 1400 K (absorbed photon fluence $\Phi = 2.3 \times 10^{13}$ cm$^{-2}$, multi-THz experiments).

The most important global output parameters of the fits are the Fermi energy $E_F = 310$ meV (obtained from the perpendicular geometry) and the diffusion time $\theta = 350$ fs (retrieved from the parallel geometry). Other parameters may vary with time after photoexcitation and with the probing electric field polarization (Fig. 5). For illustration, their steady-state values are: the carrier scattering time $\tau_s \approx 45$ fs, the carrier mobility $\eta \approx 1700$ cm$^2$V$^{-1}$s$^{-1}$, the Drude weight $D \approx 0.11$ S/ps, the plasmon resonance frequency $\omega_0/2\pi \approx 2.07$ THz and the localization parameter $c \approx -0.13$. These values compare very well with previously measured properties by using magnetoplasmon infrared spectroscopy at 4 K (which provides $E_F = 328$ meV, $\tau_s \approx 43$ fs, and $\omega_0/2\pi \approx 2.15$ THz) [45].

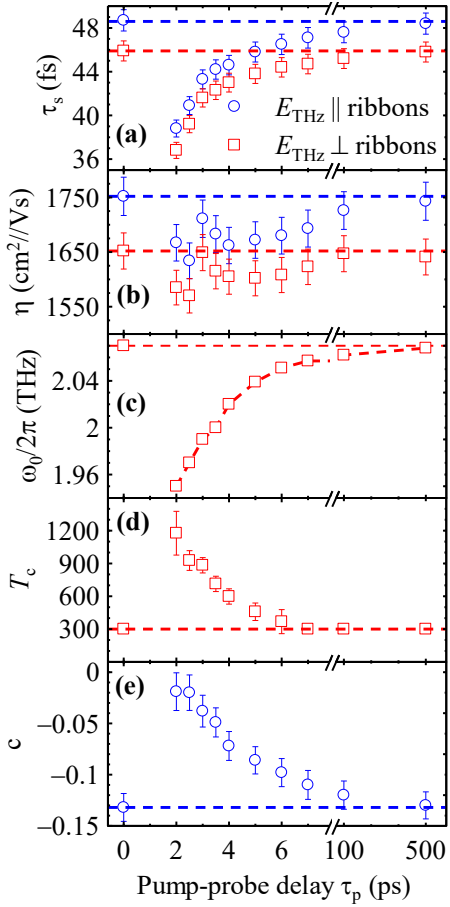

**Figure 5.** Evolution of the transport parameters as a function of the pump-probe delay deduced from the sheet photoconductivity data shown in Figs. 3 and 4 for both polarizations: (a) scattering times $\tau_s$, (b) mobility $\eta$, (c) plasmonic frequency $\omega_0$ ($E_{THz} \perp$ ribbons), (d) carrier temperature ($E_{THz} \perp$ ribbons; symbols without error bar were fixed to the room temperature in the fitting procedure), and (e) localization parameter $c$ of the modified Drude-Smith model ($E_{THz} \parallel$ ribbons).

As observed in Fig. 5, the scattering time exhibits a very small (but experimentally significant) anisotropy of $\sim 5\%$ ($\tau_{s,\perp} < \tau_{s,\parallel}$), which is probably due to an extra scattering of carriers on the ribbon edges for the probing field perpendicular to the ribbons. Consequently, the mobilities obtained for the perpendicular and parallel geometry differ by the same amount. From Fig. 5(b) we also conclude that the variation of the mobility upon photoexcitation is negligible, thus confirming the past assumption of carrier-temperature-independent mobility [8,9,24]. The mobility value obtained here is somewhat worse than the best one reported in this type of graphene [46]. This could be due to the nanofabrication-induced mobility deterioration originating from resist residua, ion contamination during the chemical processing, and rough edges created during the lithography process. Those factors may also affect the local conductivity behavior; however, they were not clearly distinguishable within our spatial resolution in Fig. 2(a).

The carrier temperature recovers with a time constant of $1.9 \pm 0.2$ ps. A similar relaxation is observed also for the carrier scattering times ($1.45 \pm 0.2$ ps) and the dynamical red shift of the plasmonic frequency ($2.2 \pm 0.4$ ps) shown in Fig. 5. The slow component observed namely in panels (a,c) is related to the cooling of the lattice and it decays within several hundreds of picoseconds [24].

The localization parameter $c$ decreases in absolute value practically to zero upon photoexcitation implying that carriers are heated enough to overcome the existing potential barriers. It then relaxes back to its steady-state value of $-0.13$ which means that the barriers start to hinder the carrier motion again as their temperature decreases. Furthermore, the characteristic localization length scale $d \approx \sqrt{10 D_{\text{dif}} \theta} \approx v_F \sqrt{5 \tau_s \theta}$ related to the diffusion time in Eq. (8) is ~300 nm. We now examine several candidates for defects which can cause such localization of carriers in graphene:

(a) Steps between terraces in the SiC substrate cause an anisotropy of dc conductivity on the level of 3% for samples with considerably higher density of the steps [47]. In our samples, the dc conductivity is expected to change by a factor of $-c \approx 13\%$ for a much lower density of the steps. Furthermore, the mean distance between the terrace steps in our sample exceeds 2 μm as seen in Fig. 2(c), which would correspond to a characteristic diffusion time of $\theta \approx 15$ ps. Such a long time would induce confinement spectral features only below 100 GHz with negligible effects in our accessible spectral range. From the point of view of both the confinement strength and confinement distance, the substrate terraces are thus not likely to be responsible for the observed confinement.

(b) Domains with different crystallographic stacking are naturally formed during the growth due to the nucleation dynamics and the built-in strain as observed in multilayer graphene [48]; their characteristic size is in the range of $\sim 0.1 - 1$ μm. However, no such features were observed in QFSLG. Instead, the large lattice mismatch between SiC and graphene leads to a surface reconstruction of QFSLG on the nanometer length scale, which is by far too small to introduce any signatures into THz spectra.

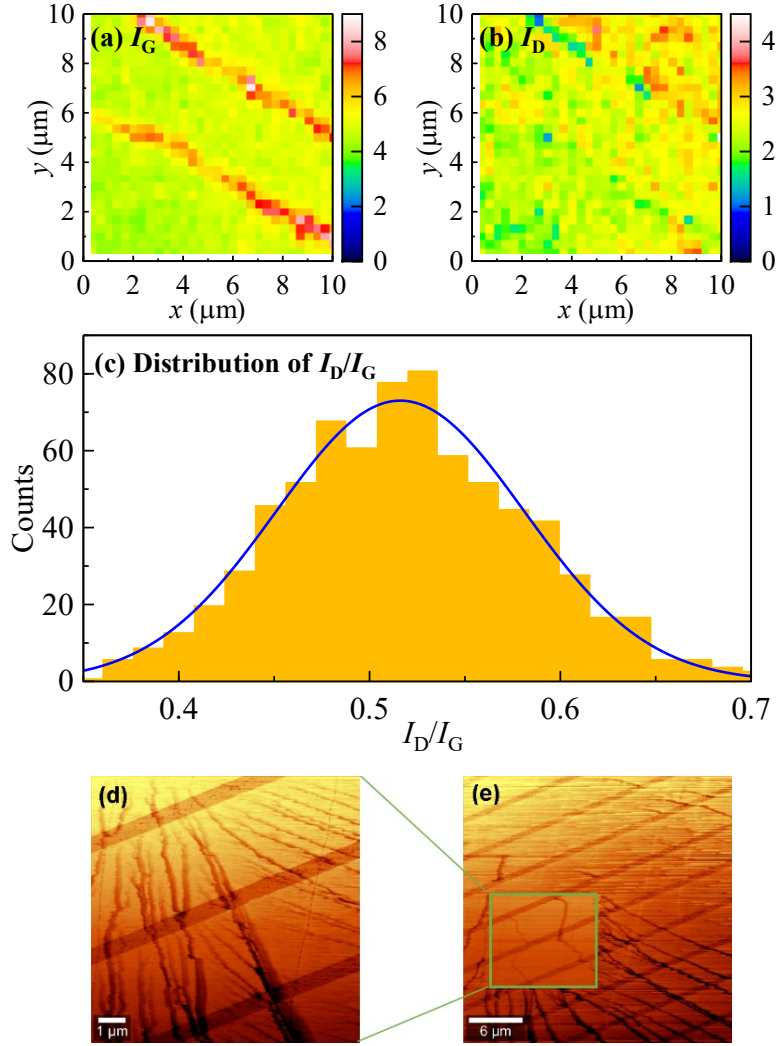

**Figure 6**. (a,b) Intensities of G and D peaks of the Raman spectra of QFSLG sample obtained prior to the lithography process. The two oblique lines in the $I_G$ spectra correspond to the terrace steps on the substrate. (c) Distribution of the intensity ratio of the D and G peaks of the Raman spectra obtained from the results in the panels (a) and (b) (data corresponding to the terrace steps were excluded) and its Gaussian fit. (d,e) AFM adhesion force scans. (d) The first scan over a smaller area shows a large density of defects. (e) Subsequent scan over a larger area shows that the first AFM scanning to a large extent cleaned the surface.

(c) In Fig. 6(a,b) we present a map of Raman intensities of the *D* and *G* lines in a $10 \times 10\ \mu m^2$ region of the sample surface measured prior to the lithography process. (The measurements were performed with Witec Alpha 300 Raman confocal microscope, laser wavelength of 532 nm, beam diameter ∼ 1 μm). The corresponding histogram of the intensity ratio $I_D/I_G$ is plotted in Fig. 6(c). It is known that this ratio correlates with the characteristic density of defects [49,50]. While the reference [50] describes the effect of point defects in the sp² carbon lattice, which are supposed to lead mainly to isotropic scattering of charge carriers and thus to a reduced mobility, Ref. [49] investigates the role of nanocrystallite boundaries, which might contribute more likely to the localization of charge carriers. From our Raman data it follows that the characteristic distance between defects is of the order of a few tens of nanometers. The carrier localization on tens of nanometers scale would correspond to the diffusion time in the order of units of femtoseconds. Such a process would be too fast to induce a dispersion in our accessible spectral range. Nevertheless, it is clear from [50] that the Raman data are quite sensitive to the short-range defects and not really to the defects separated by a few hundreds of nanometers in average. Very likely, the defects detected by Raman spectroscopy in our sample are at the origin of isotropic scattering of carriers, thus reducing the THz mobility, i.e., they may explain our reduced mobility of $1700\ \mathrm{cm^2 V^{-1} s^{-1}}$ in comparison with the value of $\sim 3000\ \mathrm{cm^2 V^{-1} s^{-1}}$ observed in the best materials [46]. The much more sparsely distributed defects introducing energy barriers and weak localization of carriers cannot be observed in the Raman scattering experiments.

(d) AFM adhesion force scan [Fig. 6(d,e)] provides information on the changes of surface properties and it can thus sense, e.g., a weak surface contamination not seen in the topographic scans. In the figure we observe a set of features across the nanoribbons with a characteristic distance of the order of hundreds of nanometers. These features do not reproduce during subsequent scans [Fig. 6(e)], which supports the view that they reflect a surface contamination upon lithographical process, which may be displaced or removed by the AFM tip [51]. However, the mechanical AFM tip cleaning [52] of the whole graphene surface for far-field THz spectroscopy is not possible due to the inherently large probed area. Although the only support is the correct characteristic distance, the surface contamination may cause weak modulation of the potential, which is now the best candidate for the explanation of the observed weak localization of charges.

### 3.3. Sub-picosecond dynamics

In order to avoid the strong artifacts appearing at the shortest pump-probe delays and to get the most complete picture of the transient data on sub-picosecond time scale, we acquired a dense two-dimensional (2D) data grid in the time domain, i.e., a series of transient THz wave forms $\Delta E(\tau, \tau_\mathrm{p})$ at closely spaced (100 fs) delays $\tau_\mathrm{p}$ between the optical pump and the THz probe. The 2D wave forms recorded for the two polarizations are shown in Figs. 7 (a) and (b) and exhibit a significant phase change with respect to each other: a rapid examination of the plots reveals that the two sets of wave forms have an opposite sign and an additional mutual time shift of their extrema by $\sim 0.16$ ps.

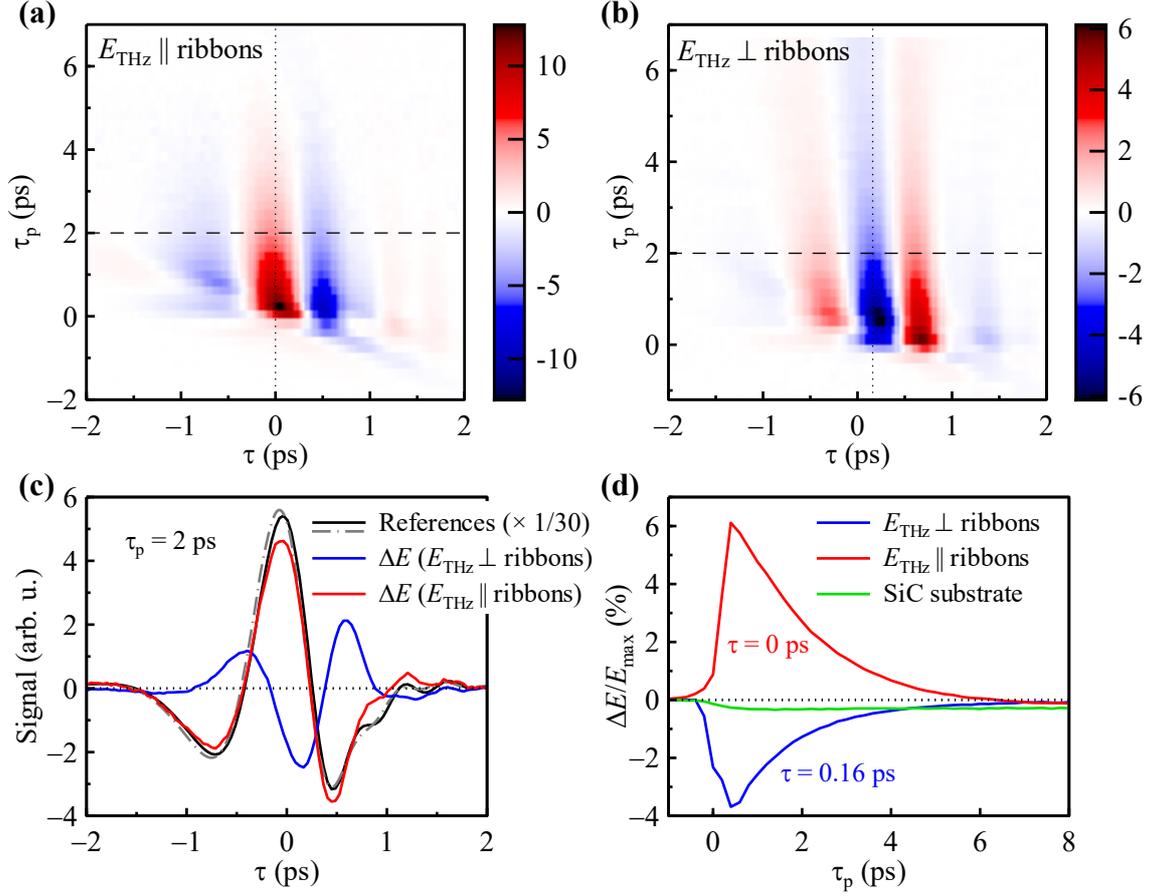

**Figure 7.** (a,b) 2D colormap plots of the transient waveforms $\Delta E(\tau, \tau_p)$ for both orientations of the ribbons. (c) Change of the THz waveform $\Delta E$ upon photoexcitation of the sample, compared to reference waveforms transmitted through an unexcited sample (solid line: parallel geometry; dash-dotted line: perpendicular geometry); pump-probe delay: $\tau_p = 2$ ps [the related cuts in 2D data are indicated by the horizontal dashed lines in panels (a) and (b)]. The mutual phase shift of the signals for parallel and perpendicular orientation is clearly visible. (d) Differential THz transmission as a function of the pump-probe delay $\tau_p$ (pump-probe scan), recorded at the highest absolute value of the differential THz waveform normalized by the highest value of the reference [the related cuts in 2D data are indicated by vertical dotted lines in panels (a) and (b)]; for comparison a differential transmission measured for the bare substrate under similar conditions is added as green line.

This difference is also illustrated in Fig. 7(c) on examples of transient wave forms obtained for a pump-probe delay of 2 ps for the two polarizations in comparison with the reference waveforms. This behavior reflects how differently the carrier excitation affects the response for the two probing polarizations. For $E_{\text{THz}} \parallel$ ribbons, the heating of excess carriers reduces the intraband conductivity (i.e., the Drude-like conductivity with a small localization contribution) of graphene in a broad spectral range; the photo-induced waveform thus has the same sign and similar shape as the reference waveform, indicating an induced transparency. The behavior for $E_{\text{THz}} \perp$ ribbons is controlled namely by a red shift of the plasmonic response upon photoexcitation; this implies a significant phase shift which yields a negative and slightly time shifted photo-induced signal. The pump-probe scans measured at the extremum of the differential waveforms shown in Fig. 7(d) thus have opposite signs.

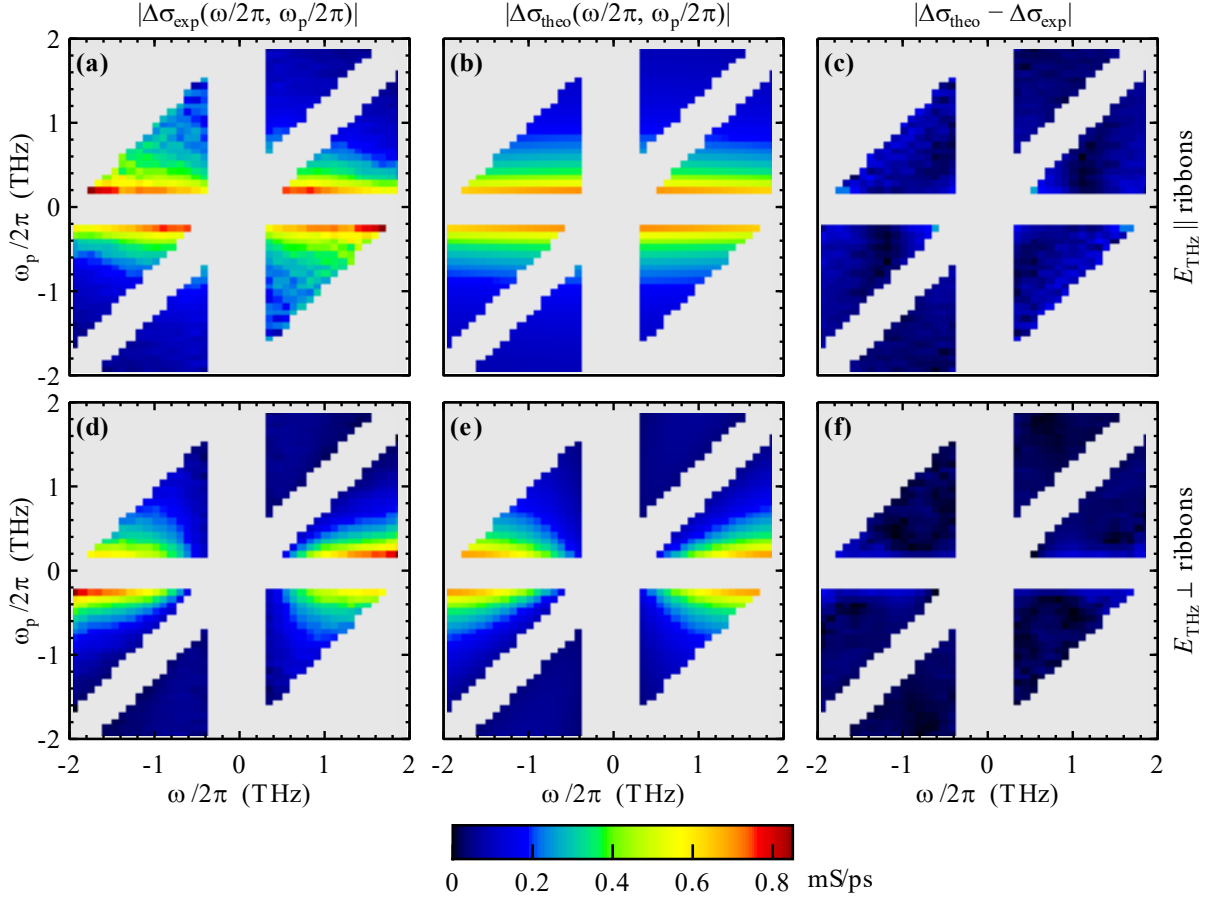

**Figure 8.** (a,d) Experimentally obtained amplitudes of the two-frequency transient sheet conductivity spectra. (b, e) The best fits of the measured data. (c,f) Amplitude of the residua of the fits (with the same color scale). Upper row: $E_{THz} \parallel$ ribbons, experimental data and a fit by the modified Drude-Smith model, Eq. (9). Lower row: $E_{THz} \perp$ ribbons, experimental data and a fit using the model where the plasmon frequency and damping change, Eq. (10). The experimentally accessible region has a complex shape due to the unavoidable frequency mixing [25]; The grey color indicates regions where at least one of the relevant measured spectra falls below the noise floor.

In order to reach sub-picosecond time resolution of the sample response function we applied the 2D Fourier transformation (apart from the Fourier transformation along the standard measurement time $\tau \to \omega$, the second transformation involves the pump-probe delay $\tau_p$; we denote the conjugated frequency by $\omega_p$) to the transient wave forms. This approach allows one to correctly account for the frequency mixing inherently present in the sub-picosecond dynamics and to deconvolute the instrumental functions [25,26,27,28].

The two-frequency transient sheet conductivity of the graphene layer can be evaluated from the previously published expressions [25] further developed in the Appendix. The resulting transient sheet conductivity spectra are shown in Fig. 8 (for simplicity we show here only the signal amplitudes; however, we stress that both the real and imaginary parts are experimentally available and that the fitting was performed on these complex spectra). The 2D spectra exhibit a center of inversion symmetry and the accessible region consists of a union of several polygons. The limits in the $\omega$ and $\omega - \omega_p$ directions are mainly due to the bandwidth of the THz pulses (the diagonal limits stem from the frequency mixing). The limits in the $\omega_p$ direction are controlled predominantly by the scan length and scan step of the pump – probe delay line through Nyquist criteria.

For $E_{\text{THz}} \parallel$ ribbons, the measured spectrum [Fig. 8(a)] resembles a pole located at $\omega = \omega_p = 0$; such a shape is compatible with the MDS model response [Fig. 8(b)] with a weak localization. The model spectrum is derived in the Appendix, Eq. (A6), and it reads:

$$\Delta\sigma_\parallel(\omega, \omega_p) = \frac{w}{L}\frac{1}{\pi}\frac{D_{\parallel,E}\dot{G}_E(\omega + i\tau_{c\parallel}^{-1}) - D_\parallel \dot{G}_G(\omega - \omega_p)}{\tau_{c\parallel}^{-1} - i\omega_p}, \quad (9)$$

where

$$\dot{G}_{G,E}(\Omega) = \frac{\tau_{s\parallel G,E}}{1 - i\Omega\tau_{s\parallel G,E}}\left(1 + \frac{c_{G,E}}{1 - i\Omega\theta_{\parallel G,E}}\right)$$

and where $\tau_{c\parallel}$ is the relaxation time from the excited state $E$ (characterized by the Drude weight $D_{\parallel,E}$, scattering time $\tau_{s\parallel E}$, diffusion time $\theta_{\parallel E}$ and localization parameter $c_E$) to the ground-state level $G$. Note that the denominator term in (9) assumes an exponential decay of the photoexcited carrier population. Upon fitting, we fixed all the ground-state parameters to the values obtained within the analysis of the slow dynamics (we also set $c_E$ to zero). The retrieved best-fit excited-state parameters are $D_{\parallel,E}$ = 0.22 S/ps, $\tau_{s\parallel E}$ = 40 fs, and $\tau_{c\parallel}$ = 2.1 ps (these parameters control the Drude peak amplitude, its width in $\omega$, and the linewidth in $\omega_p$, respectively). The relaxation time $\tau_{c\parallel}$ matches the decay time obtained from the analysis of the picosecond dynamics of the conductivity in the previous section. The (initial) scattering time $\tau_{s\parallel E}$ = 40 fs is similar to the shortest one observed in the picosecond dynamics in Fig. 5(a); this indicates that the carrier scattering is not further enhanced during the sub-ps evolution of the system after photoexcitation. The term $D_{\parallel,E}$ has the meaning of the Drude weight immediately after photoexcitation and its fitted value corresponds to an initial carrier temperature of 5000 K (Fig. 9); this temperature roughly agrees with the value estimated from the energy conservation condition at the very initial stage of the photocarrier dynamics [24]. The residua of the fit [Fig. 8(c)] provide an almost featureless image, which indicates a good quality of the fit. The time resolution here is limited by the excitation pulse width and by the step employed for scanning $\tau_p$, we thus estimate that we can resolve processes as fast as 100 fs. In other words, we can conclude that the dynamics after 100 fs and longer after photoexcitation are fully governed by the MDS response which involves a carrier relaxation and a weak progressive localization of charges; no other process is observed. The good quality of the fits of 2D data also confirms that the oscillations in Fig. 4(a) are indeed artifacts.

Note that the employed two-level picture, in which an exponential relaxation of the carrier population is assumed, is an important approximation. In reality, we should expect that all the parameters of the MDS model (including the Drude weight value) decay progressively, and, moreover, that a monotonous relaxation of the carrier temperature may lead to a non-monotonous dynamics of the Drude weight (see Fig. 9). Although the presented model should be able to capture the basic behavior, the retrieved values should be considered with reservations.

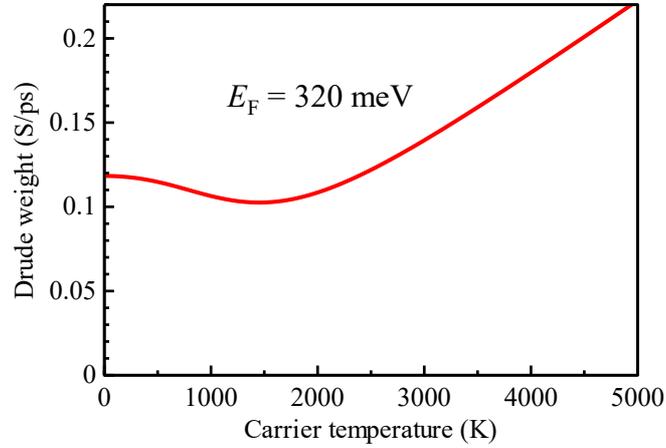

**Figure 9.** Variation of the Drude weight ($D$) versus carrier temperature ($T_c$) obtained using Eq. (5) for the Fermi energy $E_F = 320$ meV. The value of the Drude weight in photoexcited sample retrieved from the fits in 2D-frequency space for the parallel geometry ($D_{\parallel,E} = 0.22$ S/ps) agrees well with the carrier temperature of 5000 K obtained immediately after photoexcitation.

For $E_{THz} \perp$ ribbons, the measured spectrum [Fig. 8(d)] features poles located at $\omega/2\pi \approx \pm 2$ THz and $\omega_p/2\pi \approx 0$ (again, the poles are rather narrow along $\omega_p$ and broad along $\omega$), which indicates that the response of confined carriers dominates. The choice of available pertinent analytical models is very limited; we found the best match [Fig. 8(e)] with Eq. (50) in Ref. [26] which describes a pump-induced instantaneous population and a subsequent exponential depopulation of a state with different eigenfrequency $\omega_1$ and damping $\tau_{\perp 1}$ (but with the same effective Drude weight $D'_\perp$):

$$\Delta\sigma_\perp(\omega,\omega_p) = \frac{\alpha D'_\perp}{\pi} \frac{1}{X_1\left(\omega + \frac{i}{\tau_{c\perp}}\right)} \frac{\omega + \frac{i}{\tau_{c\perp}}}{\omega_p + \frac{i}{\tau_{c\perp}}} \left[1 - \frac{X_1(\omega - \omega_p)}{X_0(\omega - \omega_p)}\right] \quad (10)$$

where $X_1(\Omega) = \omega_1^2 - \Omega^2 - i\Omega/\tau_{\perp 1}$ and $X_0(\Omega) = \omega_0^2 - \Omega^2 - i\Omega/\tau_\perp$. In this model, the photoexcitation does not generate any step change in the velocity or displacement of charge carriers. [Note that this model is a considerable approximation of the plasmonic behavior following Eqs. (3) and (4) applied to ultrafast dynamics: we expect a continuous relaxation of the eigenfrequency, whereas the model (10) represents a step change in frequency of photoexcited particles and a relaxation of their population instead]. The best fit is achieved when, upon photoexcitation, the plasmon frequency changes from $\omega_0/2\pi$ = 1.94 THz to $\omega_1/2\pi$ = 1.87 THz and the damping changes from $\tau_\perp$ = 61 fs to $\tau_{\perp 1}$ = 43 fs; the population relaxation time then yields $\tau_{c\perp}$ = 1.8 ps. Also, these values match well those obtained from the analysis of the time-dependent conductivity spectra. The relaxation time matches the one obtained for the parallel orientation: $\tau_{c\perp} \approx \tau_{c\parallel}$ and it is also in agreement with the fast component inferred from the transient THz conductivity evaluated at larger pump-probe delays [between 2 and 10 ps, Fig. 4(c)] . The slight difference in the scattering times obtained from these 2D fits may be attributed to the approximations involved in the two models). The residua of the fit [Fig. 8(f)] provide an almost featureless image, therefore we can again conclude that the carrier dynamics occurring at 100 fs after photoexcitation and later is fully governed by the plasmonic relaxation, while no further process is observed.

**Conclusion**

The photoinitiated dynamics of graphene ribbons was investigated using time-resolved THz spectroscopy on a timescale 100 fs – 500 ps. Upon optical photoexcitation the carriers in graphene are heated to temperatures reaching up to 5000 K. The dynamics of the carrier temperature is a universal function governing the conductive response of the graphene ribbons within the whole investigated time interval for both parallel and perpendicular currents to the ribbon direction. The response for the probing THz electric field parallel to the ribbons has an essentially Drude-like character with signatures of a weak localization (presumably caused by contamination upon lithography); the observed photoinduced dynamics is explained in terms of a cooling and a weak progressive localization of initially free carriers. For the probing electric field perpendicular to the nanoribbons a THz plasmon develops due to the ribbon structure. The excitation decay is observed due to its frequency redshift followed on the ultrafast timescale. The nanoribbon edges weakly but significantly enhance the carrier scattering process.

**Acknowledgment**

The authors acknowledge financial support by Czech Science Foundation (grant number 24-10331S). CzechNanoLab project LM2023051 funded by MEYS CR is acknowledged for the financial support of the sample fabrication at CEITEC Nano Research Infrastructure. This work was also supported by European Union and the Czech Ministry of Education, Youth and Sports (Project TERAFIT - CZ.02.01.01/00/22_008/0004594).

**Appendix: Theory for sub-picosecond dynamics measurements**

The transient sheet conductivity in a 2D Fourier space ($\tau \to \omega$, $\tau_\text{p} \to \omega_\text{p}$) can be evaluated from a simple expression [25]

$$\Delta\sigma(\omega,\omega_\text{p}) = i\omega\varepsilon_0 \frac{T(\omega)}{\Xi(\omega,\omega_\text{p})} \frac{\Delta E(\omega,\omega_\text{p})}{E^0(\omega)} \frac{E_\text{S}(\omega)\Psi(\omega-\omega_\text{p})}{E_\text{S}(\omega-\omega_\text{p})\Psi(\omega)} \tag{A1}$$

where $T$ is the complex transmittance of the sample in the steady state, $E^0$ is the detected reference spectrum transmitted through an unexcited sample, $E_\text{S}$ is the spectrum of the THz waveform measured at the place of the sample and $\Psi$ is the spectral response function of the sensor used for this last measurement. Finally, the transfer function $\Xi$ for an ultrathin film on a substrate reads [25,26,27,28]

$$\Xi(\omega,\omega_p) = \frac{2i\omega}{c(1+n_\text{SiC})^2} \tag{A2}$$

where $n_\text{SiC}$ is the THz refractive index of the substrate (its dispersion is negligible) and $c$ is the speed of light in vacuum (not to be confused with the localization parameter in the main manuscript). Experimentally, we determined $\Delta E(\omega,\omega_\text{p})$, $E^0(\omega)$, $E_\text{S}(\omega)$, $\Psi(\omega)$ and $n_\text{SiC}$. We then used Eqs. (A1–A2) to evaluate the transient sheet conductivity $\Delta\sigma(\omega,\omega_\text{p})$ shown in Fig. 8.

Basic models for the ultrafast conductivity in two-dimensional frequency domain (Drude and oscillator models) were developed in [26]. Here we apply a similar formalism to the modified Drude-Smith response [42] and extend it to the two-dimensional frequency domain. We start from Eq. (27) in [26] which describes the time-domain response function dynamics of a system brought from the ground state $G$ to an excited state $E$. It introduces the Green's functions $G_E$ and $G_G$ describing the charge (electron or hole) trajectories in the excited and ground states, respectively. Similarly as in [26], we assume an exponential decay of the excited state population, Eq. (28) therein. We also assume that, upon photoexcitation, the system loses information about the previous particle motion due to some fast-scattering mechanism (condition of strong perturbation as introduced in [26]). For graphene stripes, where the magnitude of the charge-field interaction is described by the Drude weight, Eq. (27) in [26] leads to:

$$\Delta\sigma(\tau,\tau_p) = \frac{w}{L}\frac{1}{\pi} Y(\tau+\tau_p) \exp\left(-\frac{\tau+\tau_p}{\tau_c}\right) [D_E Y(\tau_p)\dot{G}_E(\tau) - D_G \dot{G}_G(\tau)]. \tag{A3}$$

Here $D_E$ and $D_G$ are the Drude weights in the excited and ground state, respectively, and $\tau_c$ is the population decay time. The time $\tau_p$ is the pump-probe delay and its changes are experimentally implemented by moving the delay line D2; $\tau$ has the meaning of the real time and it is implemented by moving the delay line D1 (Fig. 1) [25,26]. The expression (A3) corresponds to Eqs. (31) and (45) in [26] for the Drude and oscillator dynamics, respectively. The transformation into the 2D frequency domain reads

$$\Delta\sigma(\omega,\omega_p) = \int_{-\infty}^{\infty} d\tau \int_{-\infty}^{\infty} d\tau_p\, \Delta\sigma(\tau,\tau_p)\, e^{i(\omega\tau+\omega_p\tau_p)} \tag{A4}$$

(note that the sign convention of the imaginary part is conform to the current manuscript, i.e., it is opposite to that used in [26]).

The Heaviside step function $Y(\tau_p)$ and the definition of the Green's function $G_E(\tau)$ ensure that the term $Y(\tau_p)\dot{G}_E(\tau)$ is non-zero only when both $\tau_p, \tau > 0$. Consequently, for this term, the integration limits of both integrals in (A4) are set to 0 and $\infty$, and the expression is factorized into a $\tau$ and $\tau_p$ dependent term. On the other hand, for the $\dot{G}_G$ term, the integral reduces to the area $\int_0^\infty d\tau \int_{-\tau}^\infty d\tau_p$, which leads to a frequency mixing in the Fourier transformed Green's function. Altogether, the conductivity reads

$$\Delta\sigma(\omega,\omega_p) = i\frac{w}{L}\frac{1}{\pi}\frac{D_E \dot{G}_E(\omega + i/\tau_c) - D_G \dot{G}_G(\omega - \omega_p)}{\omega_p + i/\tau_c}. \tag{A5}$$

For the modified Drude-Smith model [42] the frequency domain Green's functions simply adopt the known form

$$\dot{G}_{G,E}(\Omega) = \frac{\tau_{G,E}}{1 - i\Omega\tau_{G,E}}\left(1 + \frac{c_{G,E}}{1 - i\Omega\theta_{G,E}}\right). \tag{A6}$$

Note that for the case of a Drude response in classical semiconductors, where $D_{E,G}/\pi \to n_0 e_0^2/m$, the ground state conductivity vanishes and no charge localization is considered, Eqs. (A5) and (A6) reduce to Eq. (33) in [26].

**References**

[1] A. N. Grigorenko, M. Polini, and K. S. Novoselov, Graphene Plasmonics, Nat. Photon. **6**, 11 (2012).
[2] F. H. L. Koppens, D. E. Chang, and F. J. García de Abajo, Graphene Plasmonics: A Platform for Strong Light–Matter Interactions, Nano Lett. **11**, 3370 (2011).
[3] T. Low and P. Avouris, Graphene Plasmonics for Terahertz to Mid-Infrared Applications, ACS Nano **8**, 1086 (2014).

[4] A. A. Dubinov, V. Y. Aleshkin, V. Mitin, T. Otsuji, and V. Ryzhii, Terahertz Surface Plasmons in Optically Pumped Graphene Structures, J. Phys.: Condens. Matter **23**, 145302 (2011).

[5] A. Yu. Nikitin, F. Guinea, F. J. García-Vidal, and L. Martín-Moreno, Edge and Waveguide Terahertz Surface Plasmon Modes in Graphene Microribbons, Phys. Rev. B **84**, 161407 (2011).

[6] W. Gao, G. Shi, Z. Jin, J. Shu, Q. Zhang, R. Vajtai, P. M. Ajayan, J. Kono, and Q. Xu, Excitation and Active Control of Propagating Surface Plasmon Polaritons in Graphene, Nano Lett. **13**, 3698 (2013).

[7] Z. Fei et al., Electronic and Plasmonic Phenomena at Graphene Grain Boundaries, Nat. Nanotechnol. **8**, 821 (2013).

[8] M. M. Jadidi, J. C. König-Otto, S. Winnerl, A. B. Sushkov, H. D. Drew, T. E. Murphy, and M. Mittendorff, Nonlinear Terahertz Absorption of Graphene Plasmons, Nano Lett. **16**, 2734 (2016).

[9] M. M. Jadidi, K. M. Daniels, R. L. Myers-Ward, D. K. Gaskill, J. C. König-Otto, S. Winnerl, A. B. Sushkov, H. D. Drew, T. E. Murphy, and M. Mittendorff, Optical Control of Plasmonic Hot Carriers in Graphene, ACS Photonics **6**, 302 (2019).

[10] J. W. Han et al., Plasmonic Terahertz Nonlinearity in Graphene Disks, Adv. Photonics Res. **3**, 2100218 (2022).

[11] X. Yao, M. Tokman, and A. Belyanin, Efficient Nonlinear Generation of THz Plasmons in Graphene and Topological Insulators, Phys. Rev. Lett. **112**, 055501 (2014).

[12] T. Guo, B. Jin, and C. Argyropoulos, Hybrid Graphene-Plasmonic Gratings to Achieve Enhanced Nonlinear Effects at Terahertz Frequencies, Phys. Rev. Appl. **11**, 024050 (2019).

[13] Y. Li et al., Nonlinear Co-Generation of Graphene Plasmons for Optoelectronic Logic Operations, Nat. Commun. **13**, 3138 (2022).

[14] X. Cai, A. B. Sushkov, M. M. Jadidi, L. O. Nyakiti, R. L. Myers-Ward, D. K. Gaskill, T. E. Murphy, M. S. Fuhrer, and H. D. Drew, Plasmon-Enhanced Terahertz Photodetection in Graphene, Nano Lett. **15**, 4295 (2015).

[15] D. A. Bandurin et al., Resonant Terahertz Detection Using Graphene Plasmons, Nat. Commun. **9**, 5392 (2018).

[16] Y. Li, P. Ferreyra, A. K. Swan, and R. Paiella, Current-Driven Terahertz Light Emission from Graphene Plasmonic Oscillations, ACS Photonics **6**, 2562 (2019).

[17] A. Singh and S. Kumar, Terahertz Photonics and Optoelectronics of Carbon-Based Nanosystems, J. Appl. Phys. **131**, 160901 (2022).

[18] B. Yao et al., Broadband Gate-Tunable Terahertz Plasmons in Graphene Heterostructures, Nat. Photon. **12**, 22 (2018).

[19] C. H. Gan, H. S. Chu, and E. P. Li, Synthesis of Highly Confined Surface Plasmon Modes with Doped Graphene Sheets in the Midinfrared and Terahertz Frequencies, Phys. Rev. B **85**, 125431 (2012).

[20] P. Q. Liu, I. J. Luxmoore, S. A. Mikhailov, N. A. Savostianova, F. Valmorra, J. Faist, and G. R. Nash, Highly Tunable Hybrid Metamaterials Employing Split-Ring Resonators Strongly Coupled to Graphene Surface Plasmons, Nat. Commun. **6**, 8969 (2015).

[21] P. Kužel and H. Němec, Terahertz Spectroscopy of Nanomaterials: A Close Look at Charge-Carrier Transport, Adv. Opt. Mater. **8**, 1900623 (2020).

[22] A. Singh and S. Kumar, Phonon Bottleneck in Temperature-Dependent Hot Carrier Relaxation in Graphene Oxide, J. Phys. Chem. C **125**, 26583 (2021).

[23] G. Wang, Y. Zhang, C. You, B. Liu, Y. Yang, H. Li, A. Cui, D. Liu, and H. Yan, Two Dimensional Materials Based Photodetectors, Infrared Physics & Technology **88**, 149 (2018).

[24] V. C. Paingad, J. Kunc, M. Rejhon, I. Rychetský, I. Mohelský, M. Orlita, and P. Kužel, Ultrafast Plasmon Thermalization in Epitaxial Graphene Probed by Time-Resolved THz Spectroscopy, Adv. Funct. Mater. **31**, 2105763 (2021).

[25] H. Němec, F. Kadlec, and P. Kužel, Methodology of an Optical Pump-Terahertz Probe Experiment: An Analytical Frequency-Domain Approach, J. Chem. Phys. **117**, 8454 (2002).

[26] H. Němec, F. Kadlec, S. Surendran, P. Kužel, and P. Jungwirth, Ultrafast Far-Infrared Dynamics Probed by Terahertz Pulses: A Frequency Domain Approach. I. Model Systems, J. Chem. Phys. **122**, 104503 (2005).

[27] H. Němec, F. Kadlec, C. Kadlec, P. Kužel, and P. Jungwirth, Ultrafast Far-Infrared Dynamics Probed by Terahertz Pulses: A Frequency-Domain Approach. II. Applications, J. Chem. Phys. **122**, 104504 (2005).

[28] P. Kužel, F. Kadlec, and H. Němec, Propagation of Terahertz Pulses in Photoexcited Media: Analytical Theory for Layered Systems, J. Chem. Phys. **127**, 024506 (2007).

[29] P. Kužel, H. Němec, F. Kadlec, and C. Kadlec, Gouy Shift Correction for Highly Accurate Refractive Index Retrieval in Time-Domain Terahertz Spectroscopy, Opt. Express **18**, 15338 (2010).

[30] M. Tinkham, Energy Gap Interpretation of Experiments on Infrared Transmission through Superconducting Films, Phys. Rev. **104**, 845 (1956).

[31] V. Skoromets, H. Němec, V. Goian, S. Kamba, and P. Kužel, Performance Comparison of Time-Domain Terahertz, Multi-Terahertz, and Fourier Transform Infrared Spectroscopies, J. Infrared Milli. Terahz. Waves **39**, 1249 (2018).

[32] H. G. Roskos, M. D. Thomson, M. Kreß, and T. Löffler, Broadband THz emission from gas plasmas induced by femtosecond optical pulses: From fundamentals to applications, Laser & Photon. Rev. **1**, 349 (2007).

[33] N. Karpowicz et al., Coherent heterodyne time-domain spectrometry covering the entire "terahertz gap", Appl. Phys. Lett. **92**, 011131 (2008).

[34] H.-K. Nienhuys and V. Sundström, Intrinsic Complications in the Analysis of Optical-Pump, Terahertz Probe Experiments, Phys. Rev. B **71**, 235110 (2005).

[35] J. Kunc, M. Rejhon, E. Belas, V. Dědič, P. Moravec, and J. Franc, Effect of Residual Gas Composition on Epitaxial Growth of Graphene on SiC, Phys. Rev. Appl. **8**, 044011 (2017).

[36] J. Kunc, M. Rejhon, and P. Hlídek, Hydrogen Intercalation of Epitaxial Graphene and Buffer Layer Probed by Mid-Infrared Absorption and Raman Spectroscopy, AIP Adv. **8**, 045015 (2018).

[37] J. Baringhaus et al., Exceptional Ballistic Transport in Epitaxial Graphene Nanoribbons, Nature **506**, 7488 (2014).

[38] A. L. Miettinen, M. S. Nevius, W. Ko, M. Kolmer, A.-P. Li, M. N. Nair, B. Kierren, L. Moreau, E. H. Conrad, and A. Tejeda, Edge States and Ballistic Transport in Zigzag Graphene Ribbons: The Role of SiC Polytypes, Phys. Rev. B **100**, 045425 (2019).

[39] W. Wang, K. Munakata, M. Rozler, and M. R. Beasley, Local Transport Measurements at Mesoscopic Length Scales Using Scanning Tunneling Potentiometry, Phys. Rev. Lett. **110,** 236802 (2013).

[40] Z. J. Krebs, W. A. Behn, S. Li, K. J. Smith, K. Watanabe, T. Taniguchi, A. Levchenko, and V. W. Brar, Imaging the Breaking of Electrostatic Dams in Graphene for Ballistic and Viscous Fluids, Science **379**, 671 (2023).

[41] A. J. Frenzel, C. H. Lui, Y. C. Shin, J. Kong, and N. Gedik, Semiconducting-to-Metallic Photoconductivity Crossover and Temperature-Dependent Drude Weight in Graphene, Phys. Rev. Lett. **113**, 056602 (2014).

[42] T. L. Cocker, D. Baillie, M. Buruma, L. V. Titova, R. D. Sydora, F. Marsiglio, and F. A. Hegmann, Microscopic Origin of the Drude-Smith Model, Phys. Rev. B **96**, 205439 (2017).

[43] T. Ostatnický, Linear THz Conductivity of Nanocrystals, Opt. Express 27, 6083 (2019).

[44] R. D. Averitt, G. Rodriguez, J. L. W. Siders, S. A. Trugman, and A. J. Taylor, Conductivity Artifacts in Optical-Pump THz-Probe Measurements of $YBa_2Cu_3O_7$, J. Opt. Soc. Am. B **17**, (2000).

[45] M. Shestopalov, V. Dědič, M. Rejhon, B. Morzhuk, J. Kunc, V. C. Paingad, P. Kužel, I. Mohelský, F. Le Mardelé, and M. Orlita, Plasmon-Plasmon Interaction and the Role of Buffer in Epitaxial Graphene Microflakes, Phys. Rev. B **108**, 045308 (2023).

[46] F. Speck, J. Jobst, F. Fromm, M. Ostler, D. Waldmann, M. Hundhausen, H. B. Weber, and Th. Seyller, The Quasi-Free-Standing Nature of Graphene on H-Saturated SiC(0001), Appl. Phys. Lett. **99**, 122106 (2011).

[47] D. Momeni Pakdehi et al., Minimum Resistance Anisotropy of Epitaxial Graphene on SiC, ACS Appl. Mater. Interfaces **10**, 6039 (2018).

[48] T. A. de Jong, E. E. Krasovskii, C. Ott, R. M. Tromp, S. J. van der Molen, and J. Jobst, Intrinsic Stacking Domains in Graphene on Silicon Carbide: A Pathway for Intercalation, Phys. Rev. Mater. **2**, 104005 (2018).

[49] L. G. Cançado, K. Takai, T. Enoki, M. Endo, Y. A. Kim, H. Mizusaki, A. Jorio, L. N. Coelho, R. Magalhães-Paniago, and M. A. Pimenta, General Equation for the Determination of the Crystallite Size La of Nanographite by Raman Spectroscopy, Appl. Phys. Lett. **88**, 163106 (2006).

[50] L. G. Cançado, A. Jorio, E. H. M. Ferreira, F. Stavale, C. A. Achete, R. B. Capaz, M. V. O. Moutinho, A. Lombardo, T. S. Kulmala, and A. C. Ferrari, Quantifying Defects in Graphene via Raman Spectroscopy at Different Excitation Energies, Nano Lett. **11**, 3190 (2011).

[51] W. Choi, M. A. Shehzad, S. Park, and Y. Seo, Influence of Removing PMMA Residues on Surface of CVD Graphene Using a Contact-Mode Atomic Force Microscope, RSC Adv. **7**, 6943 (2017).

[52] A. M. Goossens, V. E. Calado, A. Barreiro, K. Watanabe, T. Taniguchi, and L. M. K. Vandersypen, Mechanical Cleaning of Graphene, Appl. Phys. Lett. **100**, 073110 (2012).